\documentclass[12pt]{article}
\usepackage{a4wide,amssymb,latexsym,afterpage,citesort,epsfig,amsmath,multirow}
\usepackage{babel}
\def\3{\ss}
\newcommand{\Abl}{\frac{d}{d t}\,}
\newcommand{\Summe}{\sum_{i = l}^{k-1}}

\title{A hierarchical model for aging\footnotemark\footnotetext{e-mail: $^a$ug@thp.uni-koeln.de, $^b$heiko@hlrz.kfa-juelich.de, $^c$schreck@traf3.math.uni-duisburg.de}} 
\author{U.~Geppert$^{1\,a}$, H.~Rieger$^{2\,b}$, M.~Schreckenberg$^{3\,c}$\\
        {\small\it $^1$Institut f\"ur Theoretische Physik, Universit\"at
          zu K\"oln, Z\"ulpicher~Str.~77, D-50937 K\"oln}\\
        {\small\it $^2$HLRZ, Forschungszentrum J\"ulich, D-52425
          J\"ulich}\\
        {\small\it $^3$Fachbereich 11 - Mathematik,
          Gerhard-Mercator-Universit\"at, D-47048 Duisburg}}

\begin{document}
\maketitle
\begin{abstract}
We present a one dimensional model for diffusion on a hierarchical
tree structure. It is shown that this model exhibits aging
phenomena although no disorder is present. The origin of aging in this
model is therefore the hierarchical structure of phase space.
\end{abstract}

\section{Introduction}

\hskip\parindent Strongly disordered systems have been a focal point
of research in recent years. Among the most studied materials are
magnetic systems with impurities and especially spin glasses
\cite{spinglaeser}. Their dynamics can be characterized by glassy
features which are also encountered in many different systems such as
structural glasses \cite{glaeser}, polymers in random potential
\cite{polymere}, protein folding \cite{protein}, dirty superconductors
\cite{supraleiter}, charge density waves with impurities
\cite{ladungsdichtewellen} and also areas like biological evolution
\cite{Stein92}.

Spin glasses are magnetic materials with structural disorder,
e.~g.~alloys of a magnetic and a non-magnetic metal as for example
$AgMn$ and $CuMn$. An overview of different spin glass materials can
be found in \cite{spinglaeser}. In the beginning, twenty years ago,
the main interest was to find the equilibrium properties of spin
glasses, since it was expected that a physical system reaches it's
equilibrium in finite time. In spin glasses, however, the dynamics is
governed by non-equilibrium properties such as {\it aging} which was
found almost ten years later \cite{Lundgren83}.

In this context aging describes the striking effect that magnetic
properties in spin glasses depend drastically on their history (or
age) in the frozen phase. Although it is mainly referred to aging in
the context of spin glasses it can be observed in other disordered
substances \cite{agingmaterial} also. 

Up to now it is still a major task in theoretical physics to
understand the origin of aging, although many different models have
been proposed in recent years. Most of these are phenomenological
theories like the droplet model \cite{FisherHuse}, the domain growth
theory \cite{Koper} or the trap model \cite{trapmodels}. A different
approach is given by mean-field models of Ising spin glasses
\cite{meanfield} and models with a hierarchical structure in phase
space \cite{hierarchien}. Most of these models are rather successfull
in fitting experimental data although some of them differ in their
final conclusions concerning the scaling of time dependent properties.

In order to discriminate between these differing scaling assumptions
many experiments have been carried out \cite{experimente}, but the
situation remains unclear. Therefore numerical studies play an
important role as one can obtain detailed information about the
dynamical processes and the spatial correlations in glassy systems.
Much progress has been obtained recently in the numerical studies on
aging in spin glasses and other disordered systems \cite{Rieger}.
In numerical investigations the autocorrelation function can be
calculated directly and one finds a crossover from a slow
quasi-equilibrium decay for $t \ll t_w$ to a faster non-equilibrium
decay for $t \gg t_w$. The functional form of these decays, e.~g.\ for
the three dimensional Edwards-Anderson spin glass model, is algebraic
and has the scaling form 
\begin{equation}
  \label{scalinglaw}
  C (t, t_w) = t^{-x(T)} \Phi_T\left(\frac{t}{t_w}\right)\,,
\end{equation}
with $\Phi_T(y) = c_y$, for $y = 0$, and $\Phi_T(y) \propto
y^{x(T)-\lambda(T)}$, for $y \to \infty$ \cite{Rieger}. As an
important result one gets two different exponents which are
characteristic for the two dynamic regions.

Since the algebraic decay of correlation functions is also known from
diffusion models in ultrametric hierarchies
\cite{Schreck,gg-diffusion}, which can be solved analytically, and the
analogy to the structure of Parisi's symmetry breaking scheme in the
solution of the EA-model a hierarchical ansatz to explain
aging phenomena seems rather natural. In recent years
different hierarchical models for aging were proposed
\cite{hierarchien}. Most of these models incorporate randomly chosen
distributions of diffusion rates and, as a consequence, random energy
barriers and trapping times. With further assumptions concerning the
functional form of these distributions one gets aging phenomena like
the crossover from quasi- to non-equilibrium dynamics with an
algebraic scaling form according to (\ref{scalinglaw}). 

Taking long range diffusion in an ultrametric space \cite{Schreck} as
starting point one can construct a hierarchical model which shows
aging phenomena as a direct consequence of the structure and not of
random properties. This model will be presented in the next section
followed by numerical results and some conclusions.

\section{The Model}

\hskip\parindent In contrast to the known hierarchical models for
aging the presented model incorporates no kind of disorder or randomly
distributed energies. The main idea is  to take into account the
different types of dynamics found in spin glasses. At short time
scales $t \ll t_w$ one finds quasi-equilibrium dynamics and at long
times $t \gg t_w$ non-equilibrium dynamics with a crossover region for
times $t \sim t_w$. These different types of dynamics are commonly
related to regions in phase space which are accessible at different
time scales.

\begin{figure}[t]
  \begin{center}
    \leavevmode
    \epsfig{file=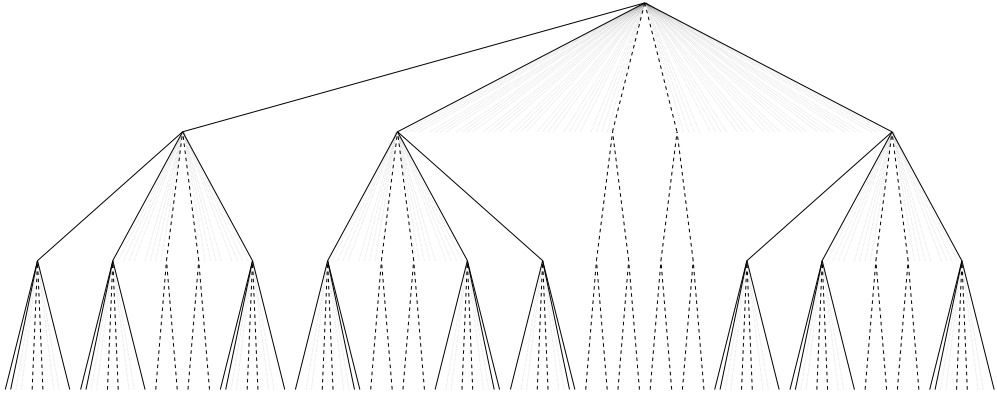}
    \caption{A hierarchical tree structure with three types of
      states characterized by different line styles: thick solid lines
      represent the surrounding non-equilibrium tree, dashed lines
      represent the quasi-equilibrium subtrees and thin solid lines
      represent the attractor region. The corresponding branching
      numbers are $n = 3, m = 2$ and $n_a = 4$; the height of the tree
      is $k = 3$.}
    \label{fig:baum}
  \end{center}
\end{figure}

A rather natural ansatz to include these different types of states is
to extend the diffusion on a hierarchical tree structure in which
every state is equal by a more sophisticated tree structure which has
at least two different types of states. The first type charactarizes
the quasi equilibrium dynamics. These states are grouped as leaves in
regularly branching trees of varying heights where standard long range
diffusion in an ultrametric space takes place \cite{Schreck}. The
second class of states is found in a regular tree structure in such a
way that the different equilibrium trees are embedded in this
surrounding tree, i.~e.\ every branching point of the non-equilibrium
tree is the root of a quasi-equilibrium subtree. The resulting
structure is shown in Figure \ref{fig:baum}, where a third class of
states is included in order to represent some kind of attractor states
for the equilibrium trees. The over-all height of the  tree is $k$,
which equals the number of different hierarchies in the tree
structure. Figure \ref{fig:baum} shows a tree of height $k=3$ and the
different trees can be recognized by the line styles, i.~e.\ the 
thick solid lines represent the surrounding non-equilibrium tree
with the branching number denoted by $n$ ($=3$), dashed lines
represent the quasi-equilibrium tree with the branching number $m$
($=2$) and thin solid lines represent the attractor states with the
branching number $n_a$ ($=4$). 

The resulting diffusion rates in the master equation, defining the
dynamics in the whole tree structure, are chosen to be unsymmetric
which takes into account the different classes represented by the
states. Let the system start at an arbitrary non-equilibrium state
$Q_0$. We then denote the probability to be found in the whole
subtree of height $l$ ($1 \leq l \leq k$) including $Q_0$ with $Q_l$,
the probability that the system can be found in
a state of an quasi-equilibrium subtree which can be reached from the
initial state by crossing $l$ hierarchies with $P_l$ and finally the
probability to be found in the attractor states corresponding to the
equilibrium subtree $P_l$ with $O_l$. The allowed long-range
hops in this tree structure are formulated in the following way
\cite{Diss}: 
%\vfill
%\newpage
\begin{enumerate}
\item From a state of the non-equilibrium type every other state
  in the tree can be reached with an rate $r_i$ with $i$ being the
  number of hierarchies between the states.
\item From an equilibrium state in $P_l$ every state within $P_l$
  can be reached with a rate $r_i$ ($1<i\leq l$) and outside of $P_l$
  every state in the tree outside from $Q_l$ is allowed with the same
  diffusion rate as for the  non-equilibrium states.
\item From the attractor states $O_l$ hops to every state in the
  tree are allowed, but with the rate $s_l$ into the corresponding
  equilibrium tree $P_l$ and with the rate $r_i$ ($l\leq i \leq k$)
  into all other states. 
\end{enumerate}
\hskip\parindent These rules include that the system has to
cross one hierarchy more in order to leave an equilibrium subtree
because the other states within $Q_l$ are excluded by the rules. This
asymmetry is chosen to generate the equilibrium subtrees $P_l$ as
dynamical traps with characteristic escape times corresponding to
their height. Using these dynamical rules the master equations can be
formulated in a closed way
\begin{eqnarray}
  \label{mastergleichungen}
  \Abl P_l(t) & = & - \Lambda_l\,P_l(t) + B_l \Omega_l(t)
              + B_l ( Q_l(t) - P_l(t) - O_l(t) )r_{l-1} + B_l
              s_l O_l(t)\,,\nonumber\\ 
  \Abl O_l(t) & = & - \Lambda_l\,O_l(t) + C_l \Omega_l(t) - B_l s_l
               O_l(t) - (A_l - B_l - C_l)r_{l-1}O_l(t)\\
              &   & + C_l ( Q_l(t) - P_l(t) - O_l(t) )r_{l-1}\,, \nonumber\\
  \Abl Q_l(t) & = & - \Lambda_l\,Q_l(t) + A_l \Omega_l(t)\,.
     \nonumber
\end{eqnarray}
To write the equations in a compact form the following abbreviations
were used:
\begin{eqnarray}
  \label{lambdaundomega}
  \Lambda_l & = & \Summe (A_{i+1} - A_{i})r_{i}\,, \\
  \Omega_l(t) & = & \Summe ( Q_{i+1}(t) - Q_i(t) - P_{i+1}(t) )r_i\,.
  \nonumber
\end{eqnarray}
The constants $A_l$, $B_l$ and $C_l$ denote the number of states in a
whole subtree $Q_l$, an equilibrium subtree $P_l$ and the attractor
states $O_l$ respectively. The diffusion rates $r_l$ and $s_l$ are as
mentioned above.

The master equations can be represented by a matrix which is an upper
triangular matrix despite some entries on the lower next-diagonal. In
principle this system can be solved for arbitrary numbers of states
and diffusion rates. In order to refer to some exact results known for
the ``classical'' diffusion models \cite{Schreck} the branching
numbers of the states are taken to be constant. Therefore the number
of states as defined above are:
\begin{eqnarray}
  A_i & = & n^i + m\frac{n^i-m^i}{n-m} +
               n_a\frac{{n_a}^i-n^i}{n_a-n}\,,\nonumber \\
  B_i & = & m^i\,,\\
  C_i & = & {n_a}^i\,;\nonumber
\end{eqnarray}
with the branching numbers $n$, $m$ and $n_a$ for the non-equilibrium,
equilibrium and attractor states respectively. If the ratio of the
diffusion rates is chosen to be constant one simply gets $r_l = r^l$
and $s_l = s^l$. With this choice the master equations are solved
numerically and some of the results are presented in the following
section. Analytic calculations concerning the exact solution of the
model can be found elsewhere \cite{Diss}.

\section{Results}

\hskip\parindent Aging phenomena are common for magnetization curves
in real spin glass materials. The corresponding function describing the
main dynamical effects for spin- or diffusion-models is the
autocorrelation function. As for magnetization experiments it is also
possible to perform waiting time dependent calculations
numerically. For such purposes one has to define the correct
waiting time dependent autocorrelation function $C(t, t_w)$ which is in
case of this hierarchical model given by 
\begin{equation}
  \label{autokorrfunktion}
  C(t, t_w ) = \sum_{l=1}^{k} P_{l0} (t) \sum_{i=l}^{k} c_{il}
  P_{l} (t_w , i) + Q_{0} (t) \sum_{i=0}^{k} c_{i0} Q_0 (t_w ,
  i) + \sum_{l=1}^{k} O_{l0} (t) \sum_{i=l}^{k} c_{il} O_{l} (t_w ,
   i)\,,
\end{equation}
where $P_{l} (t_w , i)$ is the probability that the system has reached
an equilibrium subtree of height $l$ with $i$ hierarchies between this
subtree and the starting site after the waiting time $t_w$ has
elapsed. The constant $c_{il}$ denotes the number of such subtrees to
be found. The function $P_{l0} (t)$ describes the probability that
the system is still in the same state or has returned to it after the
additional time $t$ has elapsed as it was in at the end of the waiting
time $t_w$. The probability functions of the other states are denoted
in the same way.

The waiting time dependent autocorrelation function $C(t, t_w)$ as
defined in (\ref{autokorrfunktion}) is the focal point of interest in
the following. The system parameters are the branching numbers $m$,
$n$, $n_a$ and the diffusion rates $r$ and $s$. These parameters can
be varyied in order to obtain numerical data.

\begin{figure}[p]
  \begin{center}
    %\leavevmode
    \unitlength1cm
    \begin{picture}(14.9225,7.5)
      \includegraphics{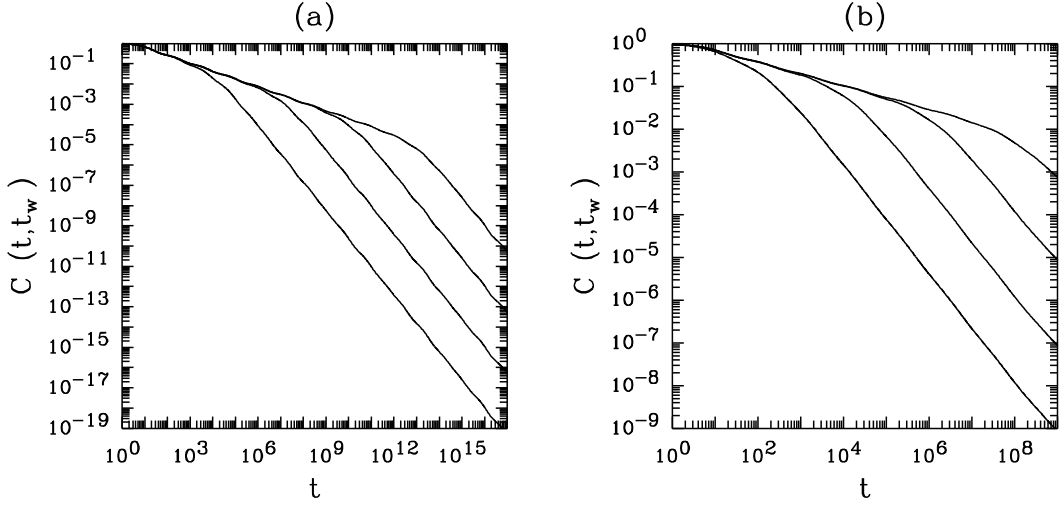}
      %\put(0,0){\framebox(14.9225,7.5)[lb]{ }}
    \end{picture}
    \caption{The waiting time dependent autocorrelation function
      $C(t,t_{\mathrm w})$ versus time $t$ for varius $t_{\mathrm w}$.
      The system parameters are as follows: $m = 3, n = 5, n_{\mathrm
        a}\ = 7, r = 0.02, s\ = 0.3;\ t_{\mathrm w} = 10^{4}, 10^{7},
      10^{10}, 10^{13}$ in (a) and $m = 2, n = 3, n_{\mathrm a}\ = 9,
      r = 0.04, s\ = 0.5;\ t_{\mathrm w} = 10^{2}, 10^{4}, 10^{6},
      10^{8}$ in (b).}
    \label{fig:autokorrmodel2}
  \end{center}
\end{figure}

\begin{figure}[p]
  \begin{center}
    %\leavevmode
    \unitlength1cm
    \begin{picture}(14.9225,7.5)
      \includegraphics{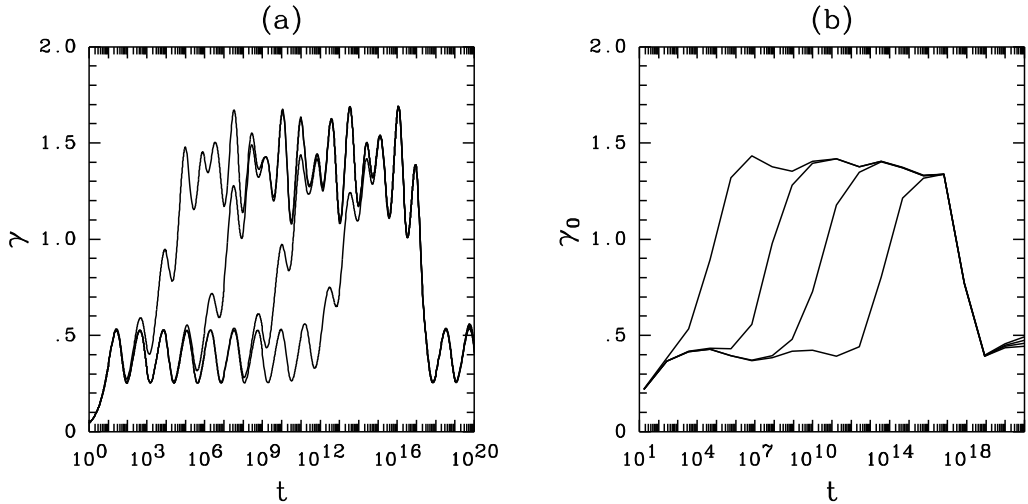}
      %\put(0,0){\framebox(14.9225,7.5)[lb]{ }}
    \end{picture}
    \caption{The exponent $\gamma$ as defined in equation
      (\ref{defexponent}) for the autocorrelation function $C(t,
      t_{\mathrm w})$ using the same parameter values as in Figure
      \ref{fig:autokorrmodel2}(a). The left diagram shows the exponent
      with the strong oscillations and the right diagram shows the
      mean value $\gamma_0$ of the exponent calculated over one period
      of oscillation.}
    \label{fig:exponentmodel2a}
  \end{center}
\end{figure}

\begin{figure}[t]
  \begin{center}
    %\leavevmode
    \unitlength1cm
    \begin{picture}(14.9225,7.5)
      \includegraphics{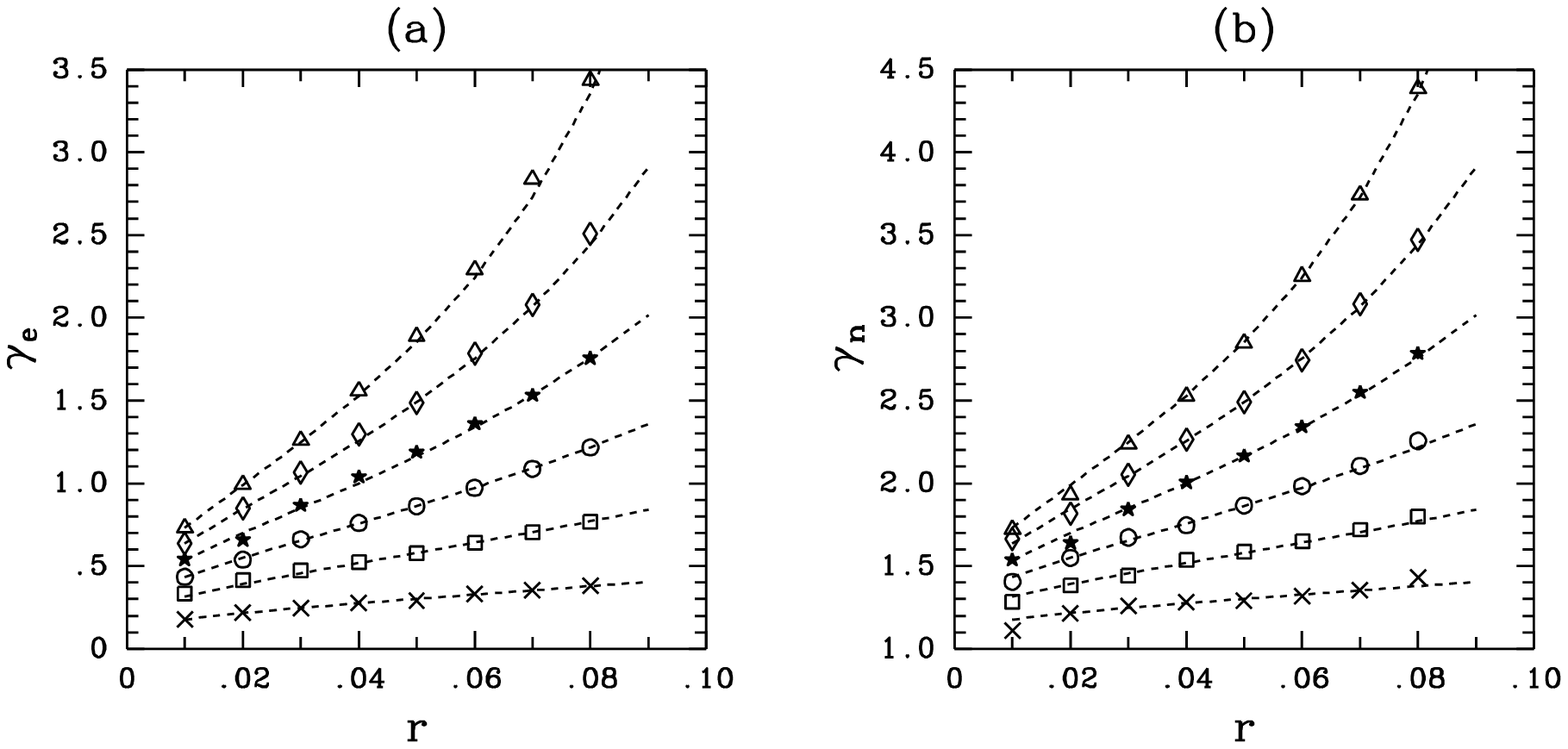}
      %\put(0,0){\framebox(14.9225,7.5)[lb]{ }}
    \end{picture}
    \caption{The exponents $\gamma_e$ and $\gamma_n$ versus the
        diffusion rate $r$ for different branching numbers $m$. The
        values of $m$ are marked with different symbols: $m = 2\
        (\times ), m = 3\ (\square ), m = 4\ (\circ ), m = 5\ (\star
        ), m = 6\ (\diamond ), m = 7\ (\vartriangle )$. The values of
        $t_w$ and $s$ vary, $n = 7$ and $n_a = 9$.}
    \label{fig:auswert1model2}
  \end{center}
\end{figure}

Figure \ref{fig:autokorrmodel2} shows the waiting time dependent
autocorrelation function $C(t,t_w)$ for parameter values as given in
the caption. A crossover from slow dynamics for small times to a
faster decay at large times charactestic for aging can be
seen very clearly. The crossover region is located at times $t \sim
t_w$. The decay obeys an algebraic time dependence although some
oscillations characteristic for self-similar systems \cite{Schreck}
occur, especially in Figure
\ref{fig:autokorrmodel2}(a). These oscillations show a strong effect
on the decay exponent which can be calculated via the logarithmic time
derivative of the autocorrelation function
\begin{equation}
  \label{defexponent}
  \gamma = - \frac{\partial \ln C(t, t_w)}{\partial \ln t}\,.
\end{equation}

The resulting exponent $\gamma$ of an algebraic decay $C(t) \propto
t^{-\gamma}$ is shown in Figure \ref{fig:exponentmodel2a}. The strong
effect of the oscillations of the autocorrelation function on the
exponent can be seen in part (a) of Figure
\ref{fig:exponentmodel2a}. As shown in part (b) the mean value
$\gamma_0$ of the exponent taken over one period is constant for the
two different time regimes $t \ll t_w$ and $t \gg t_w$. So the
autocorrelation function satisfies the scaling relation
(\ref{scalinglaw}) with temperature independent exponents $\gamma_e$
for $t \ll t_w$ and $\gamma_n$ for $t \gg t_w$. 

Since the branching numbers and the diffusion rates play an important
role in this model the exponents will at least depend on some of these
parameters. The dynamics of this model is chosen to be close to the
model solved in \cite{Schreck} so it is expected that the values of
the exponents $\gamma_e$ and $\gamma_n$ will be related to the decay
exponent calculated analytically for the simple diffusion model
\cite{Schreck}, i.~e.
\begin{equation}
  \label{exponentanal}
  \Gamma = \frac{\ln M}{| \ln R | - \ln M}\ \ \ \mbox{for}\ \ \ R <
  \frac{1}{M}\,,
\end{equation}
where $M$ is the branching number in the tree and $R$ the diffusion
rate for crossing one hierarchy. The relation $R\,M < 1$ leads to an
algebraic decay of the autocorrelation function, while one gets an
exponential decay in the other case.

The exponents $\gamma_e$ and $\gamma_n$ for different branching
numbers $m$ and diffusion rates $r$ are shown in Figure
\ref{fig:auswert1model2}. The dotted lines in diagram (a) mark the
equilibrium exponent $\Gamma$ according to equation
(\ref{exponentanal}). In the right diagram the $\gamma$-axis is
shifted by one and the lines mark the values of $\Gamma +1$. In both
diagrams the numerical values are in excellent agreement with the
theoretical prediction. As a consequence the values of $n, n_a$ and
$s$ are irrelevant for the dynamical behaviour of this model. This
aspect and it's implications will be further discussed in the next
section.

\section{Conclusion}

\hskip\parindent A diffusion model on a hierarchical tree structure
without any disorder was introduced. It was shown that the waiting
time dependent autocorrelation function $C(t, t_w)$ shows the
characteristic dependence of the waiting time $t_w$ known as
aging in spin glasses. Since the decay of the autocorrelation function
is algebraic the scaling law (\ref{scalinglaw}) is satisfied with two
constant exponents. The influence of the parameters on the two
exponents were investigated and an excellent agreement with the
equilibrium exponent $\Gamma$ was found for short times $t \ll
t_w$. After the crossover region a non-equilibrium exponent, in good
agreement with $\Gamma +1$, was found. The dynamic interpretation can
be subsumed in the following way:
\begin{itemize}
\item during the waiting time $t_w$ the system gets ``trapped'' in an
  equilibrium tree of height $l$ corresponding to a trapping time
  $\tau \sim t_w$;
\item starting the measurement of $C(t, t_w)$ after the waiting time
  causes the system to stay in this equilibrium subtree of height $l$;
\item the short time dynamics $t \ll t_w$ is therefore characterized
  by equilibrium diffusion and the equilibrium exponent $\Gamma$ is
  found;
\item in the crossover region $t \sim t_w$ the system slowly escapes
  out of this trap and the resulting exponent rises;
\item the long time behaviour $t \gg t_w$ corresponds to
  non-equilibrium dynamics with an exponent $\Gamma +1$;
\item the attractor states $O_l$ together with a large diffusion
  rate $s$ forces the system into the equilibrium subtrees;
\item since the system can be found in states of the equilibrium
  subtrees most of the times the dynamics is completely governed by
  the branching number of the equilibrium states.
\end{itemize}

The presented model includes no kind of disorder but solely a
hierarchical structure of phase space so it is obvious that the
hierarchical organization of traps or valleys in the energy-landscape
plays an essential role in explaining the aging phenomena. The results
presented here are in good agreement with the scaling assumption
(\ref{scalinglaw}) and coincide with the results of other hierarchical
models for aging \cite{hierarchien} and numerical investigations
\cite{Rieger}.

Since the world of spin glasses and aging is much richer than
discussed here, there are of course some additional phenomena which
should be examined. In relation to experimental results one should
consider temperature steps or more sophisticated temperature
cycles. The model defined above is not directly temperature
dependent, but with a straight forward relation between temperature
and diffusion rates one can map temperature steps on corresponding
steps in the diffusion rates \cite{Diss}. This should be the focus for
further investigations. 

{\bf Acknowledgements:} 
This work has been performed within the research program of the
Sonderforschungsbereich 341 (K\"oln-Aachen-J\"ulich). HR's work 
has been supported by the Deutsche Forschungsgemeinschaft (DFG).

%%%%%%%%%%%%%%%%%%%%%%%%%%%%%%%%%%%%%%%%%%%%%%%%%%%%%%%%%%%%%%%%%%%%%%%
%
%       Literaturliste
%
%%%%%%%%%%%%%%%%%%%%%%%%%%%%%%%%%%%%%%%%%%%%%%%%%%%%%%%%%%%%%%%%%%%%%%%

\end{document}